\documentclass[pdflatex,sn-nature]{sn-jnl}

\usepackage{graphicx}%
\usepackage{multirow}%
\usepackage{amsmath,amssymb,amsfonts}%
\usepackage{amsthm}%
\usepackage{mathrsfs}%
\usepackage[title]{appendix}%
\usepackage{xcolor}%
\usepackage{textcomp}%
\usepackage{manyfoot}%
\usepackage{booktabs}%
\usepackage{algorithm}%
\usepackage{algorithmicx}%
\usepackage{algpseudocode}%
\usepackage{listings}%
\usepackage{caption}

\theoremstyle{thmstyletwo}%

\theoremstyle{thmstylethree}%

\begin{document}

\title[Economy and Geography Shape the Collective Attention of Cities]{Economy and Geography Shape the Collective Attention of Cities}


\author*[1]{\fnm{Ke-ke} \sur{Shang}}\email{kekeshang@nju.edu.cn}
\equalcont{These authors contributed equally to this work.}

\author[1]{\fnm{Jiangli} \sur{Zhu}}\email{zhujl@nju.edu.cn}
\equalcont{These authors contributed equally to this work.}

\author[1,2]{\fnm{Junfan} \sur{Yi}}\email{junfan.yi@research.uwa.edu.au}
\equalcont{These authors contributed equally to this work.}

\author[1]{\fnm{Liwen} \sur{Zhang}}\email{liwenzhang@smail.nju.edu.cn}

\author[3,4]{\fnm{Junjie} \sur{Yang}}
\email{junjie.yang@yale.edu}

\author[1]{\fnm{Ge} \sur{Guo}}\email{guoge0820@hotmail.com}

\author[1]{\fnm{Zixuan} \sur{Jin}}\email{zixuanjin@smail.nju.edu.cn}

\author*[2]{\fnm{Michael} \sur{Small}}\email{michael.small@uwa.edu.au}

\affil*[1]{\orgdiv{Computational Communication Collaboratory}, \orgname{Nanjing University}, \orgaddress{Nanjing University, Nanjing, 210093, China}}

\affil*[2]{\orgdiv{Complex Systems Group, Department of Mathematics and Statistics}, \orgname{The University of Western Australia}, \orgaddress{Crawley, Western Australia, 6009, Australia}}

\affil[3]{\orgdiv{Department of Philosophy}, \orgname{Yale University}, \orgaddress{New Haven, 06511, USA}}

\affil[4]{\orgdiv{Department of Philosophy}, \orgname{Peking University}, \orgaddress{Beijing, 100871, China}}


\abstract{Complex networks are commonly used to explore human behavior. However, previous studies largely overlooked the geographical and economic factors embedded in collective attention. To address this, we construct attention networks from time-series data for the United States and China, each a key economic power in the West and the East, respectively. We reveal a strong macroscale correlation between urban attention and Gross Domestic Product (GDP). At the mesoscale, community detection of attention networks shows that high-GDP cities consistently act as core nodes within their communities and occupy strategic geographic positions. At the microscale, structural hole theory identifies these cities as key connectors between communities, with influence proportional to economic output. Overlapping community detection further reveals tightly connected urban clusters, prompting us to introduce geographic and topic-based metrics, which show that closely linked cities are spatially proximate and topically coherent. Of course, not all patterns were consistent across regions. A notable distinction emerged in the relationship between population size and urban attention, which was evident in the United States but absent in China. Building on these insights, we integrate key variables reflecting GDP, geography, and scenic resources into regression model to cross-verify the influence of economic and geographic factors on collective user attention, and unexpectedly discover that a composite index of population, access, and scenery fails to account for cross-city variations in attention. Our study bridges the gap between economic prosperity and geographic centrality in shaping urban attention landscapes.}




\maketitle

Herbert Alexander Simon introduces the concept of attention economics, highlighting that with growing information abundance, human attention constitutes a scarce resource \cite{simon1973applying}. This foundational insight has given rise to the study of collective attention, where large groups of users simultaneously focus on specific digital content, revealing competitive dynamics that shape online engagement and information diffusion \cite{wu2007novelty,cattuto2009collective}.
On the other hand, social scientists have increasingly acknowledged the potential and significance of applying complex network theory in urban planning, which is utilized for traffic flow prediction and population pattern recognition\cite{wang2024role}. With the groundbreaking achievements of complex networks\cite{barabasi1999emergence,barabasi2005origin}, an emerging research trend seeks to adopt attention networks, which were originally developed to model the dynamics of collective attention\cite{wu2007novelty,cattuto2009collective,lehmann2012dynamical}, to elucidate issues related to urban development and human activities\cite{Eubank2004ModellingDO,kontokosta2024socio}, addressing the complexity and uncertainty inherent in urban systems. 
The fundamental role of attention networks in understanding how urban identities are constructed and internalized within collective human cognition is increasingly recognized. However, while current research primarily revolves around the impact of attention networks on urban morphology and mechanisms \cite{zhao2024unravelling}, the cognitive and cultural foundations of the urban attention component within these networks remain underexplored.
This gap calls for a focus on the very nature of the \textit{city genome} embedded in public perception.

Moreover, community detection focuses on identifying cohesive groups within complex networks\cite{girvan2002community}. Communities, fundamental substructures in network science, are formally defined as groups with significantly denser internal edges than external ones \cite{shang2020novel}. Rooted in the tenets of community detection, in the interconnected digital realm, recent studies observe the dynamics of collective attention, particularly in the tourism domain, intricately intertwined with the geographical and economic factors of cities\cite{xu2022tourism,kim2020exploring}. Following those perspectives, we mainly explore attention networks utilizing community detection, complemented by conventional economic methodologies for cross-validation. 

We construct attention networks using dynamic shifts in public focus during major lockdown periods in the United States and China, two leading economies with distinct cultural contexts, when restricted movement drove digital engagement and the easing of restrictions triggered a sharp rise in travel-related attention (Extended Data \autoref{fig:ml}), revealing an intrinsic and deeply rooted collective attention toward urban destinations. Specifically, user-generated content and behavioral data from TripAdvisor and Ctrip serve as high-fidelity proxies for collective attention, capturing public interest in the United States and China, respectively. Each node in those networks represents a city, and a link between two nodes is established when users mention both cities in their travel itineraries, indicating the flow of user attention. The emergent complex system, formed by the aggregation of all such user-generated connections, reflects the dynamics of collective attention. The attention network in the United States comprises $396$ nodes and $1865$ edges, and in China, it consists of $128$ nodes and $461$ edges (Extended Data \autoref{cc}). To strengthen our analysis and bridge the domains of network science and computational economics, we incorporated comprehensive economic and geographical data from various official sources. Economic and demographic statistics were sourced from the United States Bureau of Economic Analysis, and the National Bureau of Statistics of China. Geographical features, including administrative boundaries, railway stations, and airports, were obtained from 12306 China Railway, OpenStreetMap, along with official government repositories (Extended Data \autoref{data_sources}).

\section{Uncovering GDP footprint in collective attention}
\begin{figure}
\centering
\includegraphics[width=\textwidth]{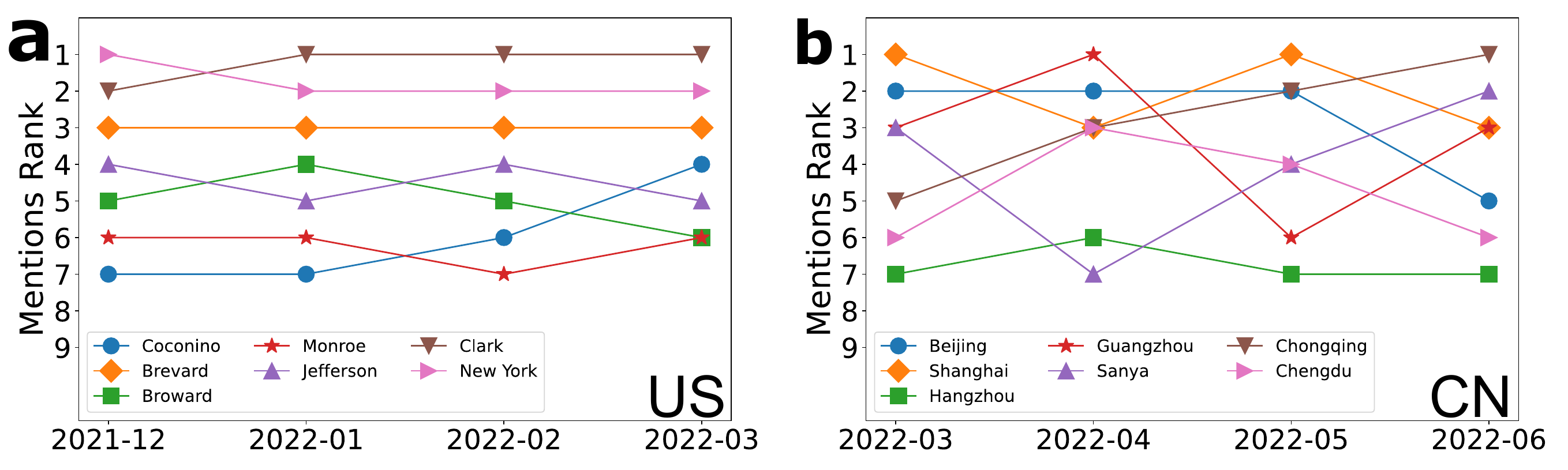}
\caption{\label{Ranking} The ranking changes of cities in the top ten for consecutive four months. Each time a city is mentioned by a user post, it counted as one mention. The cities are then ranked based on their monthly mention counts. Across the United States and China, a consistent presence of seven cities each has been observed in the rankings. To ensure a comparable network scale, we use counties in the United States and prefecture-level cities in China as the geographic units of analysis. For consistency, we refer to both types of administrative units as cities. All cities analyzed in this paper are located within the contiguous United States and mainland China.}
\end{figure}

In assessing online social attention, the volume of referenced content stands as a natural metric for collective attention. From a macro-perspective, we examine the attention rankings of various cities (\autoref{Ranking}) and find that economically dominant metropolitan areas consistently rank among the top ten in collective user attention. Notable examples from the United States include New York County, which encompasses the economic core of Manhattan, and Clark County, anchored by Las Vegas. In China, Beijing serves as the political center and is also one of the most economically powerful cities, while Shanghai stands out as a major financial hub.

Given that city classifications typically incorporate both economic output and population size as key criteria, we further investigate the relative influence of these factors on online visibility. To evaluate the independent contributions of GDP and population to the number of city mentions, we conduct a multiple linear regression analysis (LR) \cite{galton1886regression}. This model allows us to assess how each factor, GDP and population, affects city mentions while controlling for the other. Following standard econometric practice, T-tests were applied to assess the statistical significance of the coefficients. As shown in \autoref{tab:city_data}, the results reveal a robust positive association between GDP and city mention counts across major urban centers in both countries. However, the population shows a statistically significant positive correlation with the frequency of mention in the United States, but no such association is observed in China. This contrast highlights a clear divergence between Western and Eastern contexts, and in particular, as a populous nation with numerous cities exceeding one million residents, the sheer abundance of large urban centers in China may dilute public attention toward population size as a distinguishing factor.

\begin{table}[htbp]
\centering
\caption{Multiple Linear Regression of 2022 city mention counts on 2021 GDP and population. 2021 GDP, available one year before the observation period, is used as a lagged predictor. This aligns with standard economic practice for modeling the delayed impact of economic conditions on collective attention. The coefficient of determination ($R^2$) measures the proportion of variance in mention frequency explained by the model. All $p$-values are two-sided, assessing statistical significance in either direction. In economics, significance is commonly evaluated at the $10\%$, $5\%$, and $1\%$ levels ($p < 0.1$, $p < 0.05$, $p < 0.01$). For Chinese cities, the model exhibits very strong explanatory power ($R^2 = 0.614$), driven entirely by GDP, which is highly significant ($p < 0.001$), while population is not statistically significant ($p = 0.593$). For the United States, an $R^2$ of $0.270$ is considered substantial in studies of human behavior, where models often face limitations due to unobserved variables, and both GDP and population are highly significant predictors ($p < 0.001$).}
\label{tab:city_data}
\begin{tabular*}{\textwidth}{@{\extracolsep{\fill}} l c c c }
\toprule
Country & $R^2$ & GDP ($p$\text{-}value) & Population ($p$\text{-}value) \\
\midrule
United States & 0.270 & $< 0.001$ & $< 0.001$ \\
China         & 0.614 & $< 0.001$ & 0.593 \\
\bottomrule
\end{tabular*}
\end{table}

\begin{figure}[h]
  \centering
  \includegraphics[width=1\textwidth]{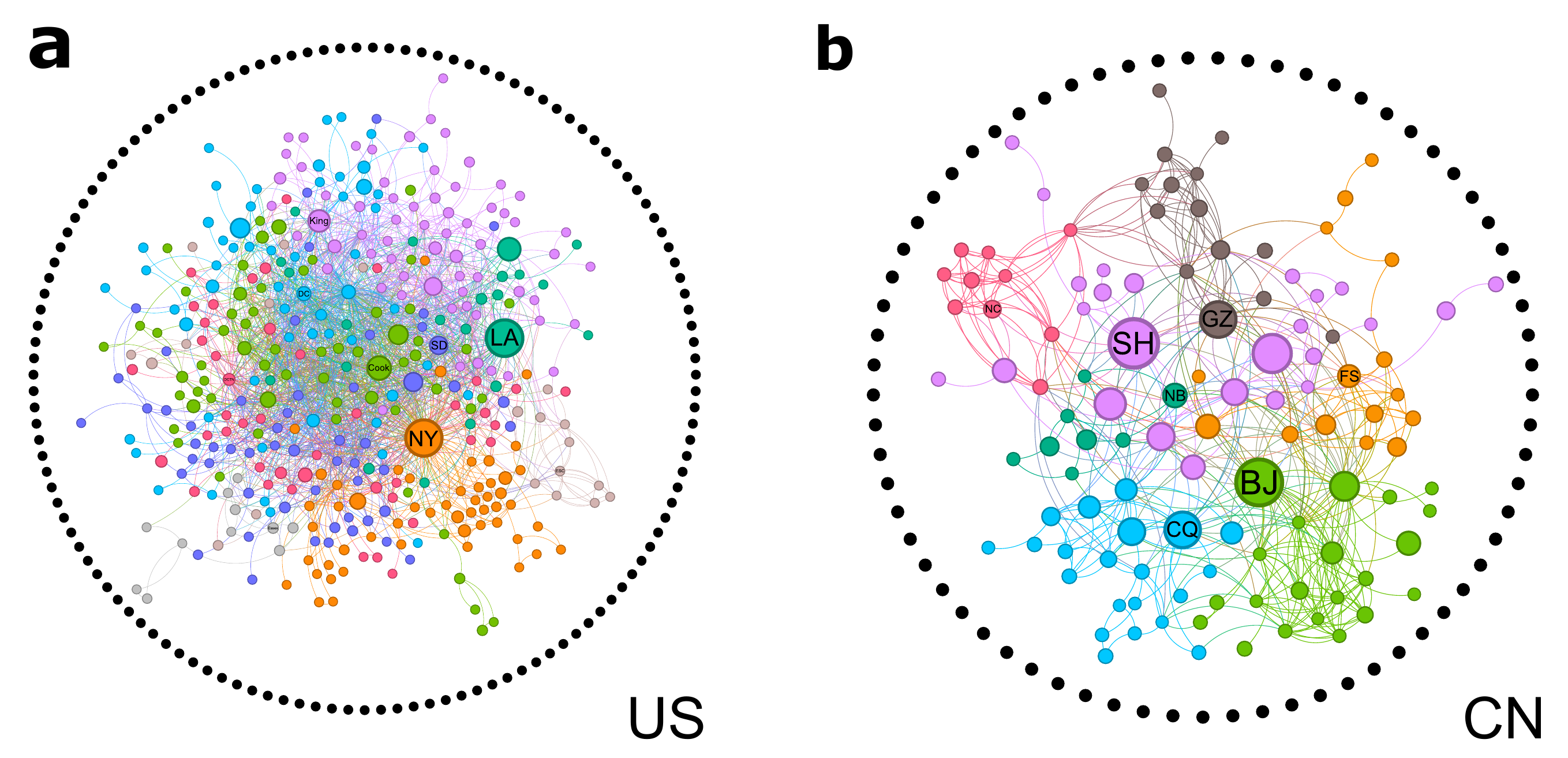} \hfill
  \caption{City attention networks based on TripAdvisor and Ctrip data, with communities divided by Louvin algorithm. Node sizes are proportional to node degree, and colors represent different major communities. In the United States network (a), major hubs such as Los Angeles, New York, Cook County, and King County emerge as high-degree nodes within their respective communities, each situated in a key economic region across the contiguous United States. In the Chinese network (b), Shanghai, Beijing, Guangzhou, and Chongqing emerge as central nodes within their respective communities, and correspondingly represent the major economic centers in eastern, northern, southern, and western China.}
  \label{com}
\end{figure}

At the mesoscale, we reveal urban agglomerations in collective attention networks via the community detection theory. These network substructures closely reflect the empirically observed economic and geographic regional structures (\autoref{com}). Here, we compare four widely used algorithms, the Label Propagation Algorithm (LPA) \cite{raghavan2007near}, the Laplacian Dynamics (LD) \cite{lambiotte2014random}, the Louvain \cite{blondel2008fast} and the Girvan-Newman (GN) \cite{girvan2002community}, to identify the optimal community structure based on the default metric Modularity (Q) \cite{newman2004finding}, and find that the Louvain method achieves the highest modularity across all countries (Supplementary Information \autoref{cda}). This is consistent with our previous demonstration that the Louvain algorithm exhibits a strong performance in detecting ground-truth communities, even compared to state-of-the-art embedding and deep learning methods \cite{ran2025machine}.
As depicted in \autoref{com}, within these detected communities, central nodes with high degree (Extended data \autoref{cen}) are predominantly high-GDP cities in the United States and China, highlighting their critical roles as economic hubs within their respective regional clusters. In the United States, New York in the Northeast anchors the financial and service core, Los Angeles in the West leads the innovation and creative economy, Chicago in the Midwest forms a central hub for urban connectivity, and San Francisco in the Western Coast stands as a major technology center. In China, Beijing in the north serves as the political and cultural center, Shanghai on the eastern coast drives the Yangtze River Delta economy, Guangzhou in the south represents a major southern commercial and transportation hub, and Chongqing in the west functions as a critical economic engine for inland development. 

Subsequently, for each detected community, the total mention count was regressed on total GDP. 
A strong positive association was observed in both the United States ($R^2 = 0.906$, $p < 0.001$) and China ($R^2 = 0.976$, $p < 0.001$), indicating that communities with higher economic output attract significantly greater collective attention. 
The exceptional explanatory power of GDP at the meso-level suggests that economic output extends beyond the aggregated perceptions of all cities and becomes constitutive of collective attention toward urban agglomerations. 

Building on the mesoscale community structure, where urban group attention shows a strong association with regional GDP, we extend the analysis to the microscale network topology to examine the strategic positioning of individual cities. The attention network displays both direct linkages and structural discontinuities among communities, indicating heterogeneous pathways for information diffusion. To quantify the capacity of a city to access diverse information flows, with attention flow serving as their empirically observable realization, we adopt the structural hole index known as effective size \cite{burt2004structural}, which measures the extent to which a node maintains non-redundant connections across communities. A higher effective size reflects greater potential for brokering attention information among otherwise disconnected groups, thus indicating the influence of a city within the network (Supplementary information \autoref{ef}). Linear regression reveals a significant positive correlation between effective size and city-level GDP in both the United States ($R^2 = 0.299$, $p < 0.001$) and China ($R^2 = 0.366$, $p < 0.001$). Our finding illustrates that cities with higher economic output shape regional clusters and simultaneously occupy central brokerage roles at the micro-level, reinforcing the view that economic output is systematically embedded in the architecture of collective attention across scales.

Taken together, our analysis across macro-, meso-, and microscales demonstrates that economic output is deeply embedded in the structure of collective attention toward cities. At the microscale, high-GDP cities occupy central brokerage roles; at the mesoscale, community-level GDP explains over $90\%$ of the variance in collective attention within urban agglomerations;  and at the macroscale, national attention dynamics align closely with aggregate economic trends.  The exceptional explanatory power at the mesoscale highlights the role of GDP not merely as a correlate, but as a constitutive socioeconomic signal. GDP thus emerges as a footprint of collective attention, systematically shaping how cities are perceived in the digital public sphere.

\section{Identifying geographic elements in collective attention}
Our investigation of the economic organization within communities, specifically the role of GDP in shaping internal attention hierarchies, reveals an unexpected geographical pattern, the most central cities across different communities are systematically positioned in distinct geographical directions, such as the eastern, western or northern regions of major nations including the United States and China (\autoref{com}). Urban agglomerations inherently overlap in geographical and cultural space, which demands analytical frameworks that transcend discrete community boundaries. To address this, we propose a novel framework for overlapping community detection, integrating geospatial coordinates and cultural dimensions. This approach reveals how economic strength, network proximity, and shared cultural context converge within the structure of urban attention.  

Central to our framework is the K-clique Percolation Method (CPM) \cite{Palla2005UncoveringTO}, chosen for its ability to identify overlapping communities through local cohesion. CPM identifies communities as percolating clusters of fully connected K-cliques. A K-clique is a complete subgraph where every city is linked to every other. It represents the most fundamental unit of intense urban cohesion, where economic, cultural, and infrastructural ties reach their peak density (see Supplementary Information 5). The involvement of cities in multiple K-cliques captures both their integration within tightly connected clusters and their roles within a dynamic, large-scale structure. 

\begin{figure}
  \centering
  \includegraphics[width=1\textwidth]{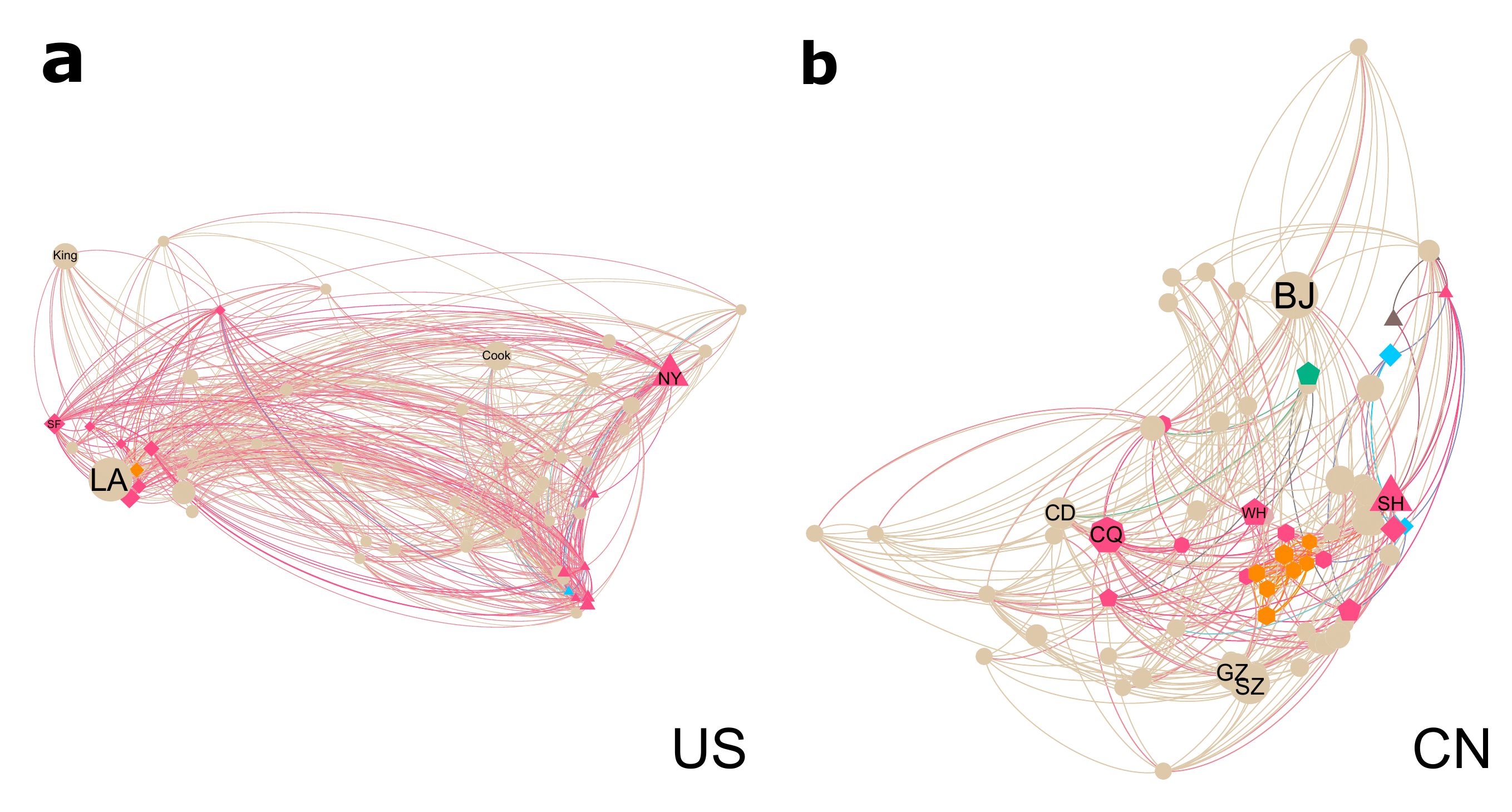} \hfill
  \caption{Overlapping community structure in attention networks. Red nodes belong to multiple communities. Node shapes correspond to their primary community, determined by the community in which the node has the highest degree rank. Panel a shows the United States with a nationwide community centered on New York and with Los Angeles as a secondary hub, reflecting a core–periphery structure at the national scale. New York also anchors a dense East Coast cluster, while San Francisco leads a distinct West Coast community. Panel b shows China with a dominant nationwide community led by Beijing, encompassing major cities such as Shanghai and Guangzhou from all regions, connected by extensive transport links. We also clearly observe an overlapping community led by Shanghai, extending in an east-west direction and connecting major central cities such as Wuhan and the western municipality of Chongqing.}
  \label{overcom}
\end{figure}

As illustrated in \autoref{overcom}, the CPM reveals a closely interconnected and unique overlapping community structure in the attention networks of both the United States and China, providing a basis for examining the geographic compactness of urban agglomerations. In the United States, a geographically extensive community emerges through the east-west coordination between New York and Los Angeles, linking major population centers across the continent and reflecting a transcontinental integration of attention flows. Meanwhile, New York forms a densely interconnected cluster along the East Coast, integrating major metropolitan areas from the Northeast to the Southeast and underscoring its central role in national-scale connectivity. In contrast, San Francisco organizes a cohesive regional community on the West Coast, reflecting strong local integration, though largely confined to the western part of the country. These high-density groupings illustrate a structure operating at multiple levels, in which key cities simultaneously participate in both regional clusters and the broader national network, enabling integrated attention flows across geographic scales. 
In China, the network exhibits a more polycentric and regionally segmented architecture. One is a national-scale community centered on Beijing, covering major cities across the country. The other is an overlapping community extends westward from Shanghai, bridging key inland hubs such as Wuhan and Chongqing, forming a prominent east–west corridor that connects coastal dynamism with inland development zones. Localized groupings also appear, such as cities around Shanghai and a coastal set of smaller cities including Ningbo, Zhoushan and Yantai. These groups have dense internal connections, revealing a clear geographic clustering. This structure underscores the role of major cities as integrative nodes across multiple overlapping subsystems, balancing national cohesion with regional specialization.

With CPM applied to detect overlapping communities based on the most tightly connected k-cliques, we observe an initial pattern of geographic proximity among certain urban groups. This provides the basis for assessing spatial coherence through an analysis of the geographical compactness of constituent cities. Consequently, we propose a geographical metric (GM), derived from the principle of Jaccard similarity \cite{jaccard1901etude} and adapted to continuous spatial coordinates, which compares the local spatial density of a community to the global distribution of all cities. Specifically, the variances of longitude and latitude in a local urban agglomeration (detected community) should be smaller than those in global cities.

\begin{equation}
    \mathrm{GM}_i = \frac{D_i}{D_{\mathrm{all}}},
    \label{eq:gm}
\end{equation}
where $D_i$ denotes the spatial extent of community~$i$, approximated by the product of the standard deviations of longitude ($s_{\mathrm{lon},i}$) and latitude ($s_{\mathrm{lat},i}$) of its constituent cities, and is given by $D_i = s_{\mathrm{lon},i} \times s_{\mathrm{lat},i}$. $D_{\mathrm{all}}$ represents the spatial extent of all cities across the entire network, computed using the overall standard deviations $\sigma_{\mathrm{lon}}$ and $\sigma_{\mathrm{lat}}$, and is given by $D_{\mathrm{all}} = \sigma_{\mathrm{lon}} \times \sigma_{\mathrm{lat}}$. A $\mathrm{GM}_i$ value close to $1$ indicates that the community occupies a spatial scale comparable to the overall network, while values much smaller than $1$ suggest strong geographic concentration.
A $\mathrm{GM}_i < 1$ indicates that community~$i$ is more geographically concentrated than the network average, suggesting a strong spatial coherence and the successful identification of a geographically localized urban cluster.

Based on a similar principle, we extend the assessment to topic coherence by introducing the community relative coherence (CRC) for each urban cluster. Using the robust Latent Dirichlet Allocation (LDA) model~\cite{Blei2003} to identify topics, we calculate their coherence using the Coherence Value (CV) metric (Supplementary Information 5). For community~$i$, we define CRC as the ratio of its internal topic coherence to that of the entire network.

\begin{equation}
    \mathrm{CRC}_i = \frac{\mathrm{CV}_i}{\mathrm{CV}_\mathrm{all}},
    \label{eq:crc}
\end{equation}
where $CV_i$ denotes the topic coherence of community~$i$, and $CV_{all}$ denotes the topic coherence of all communities. 
A $\mathrm{CRC}_i > 1$ indicates that the community exhibits stronger topic consistency than the network average, indicating a shared cultural orientation and more aligned patterns of attention within the community.

Our results show that urban attention networks in both the United States and China reveal structured patterns of spatial and topical clustering despite differences in network configuration. In the United States, one community spans a broad geographic extent with a geographic metric (GM) near $1$ indicating a spatial spread comparable to the overall network. Two more communities however exhibit much tighter concentration with GM values of approximately $0.16$ and sizes of $8$ cities each reflecting regionally localized groupings. All three US communities show a Community Relative Coherence (CRC) greater than $1$ meaning their internal topic coherence exceeds the network average. In China, $6$ communities are identified ranging in size from $4$ to $56$ cities. The largest community has a GM of $0.85$ suggesting a wide but still bounded spatial coverage. The remaining communities are markedly more compact with GM values as low as $0.058$ and several below $0.2$ indicating strong local clustering particularly among coastal and regional urban clusters. Despite their varying sizes, all $6$ Chinese communities also exhibit CRC greater than $1$, demonstrating higher internal topic coherence than the network as a whole.

These findings reveal a consistent organizational principle where collective attention forms concentrated clusters across space and topics. Spatial factors are deeply embedded in human attention, as evidenced by the consistent co-occurrence of geographic proximity and topic coherence across both countries. Even in digital networks, location continues to shape what people pay attention to.

\section{Empirical validation in economics}\label{eco}
The patterns revealed by the complex network analysis indicate that urban visibility is shaped by local endowments and systemic connectivity within the national tourism system. To quantitatively assess these insights, a computational socioeconomics approach is adopted that integrates large-scale digital trace data with curated socioeconomic indicators. Our framework enables an independent evaluation of how intrinsic appeal and regional integration influence urban prominence, with Tourism Quality (TQ), Travel Competitiveness (TC), Transportation Potential (TP), and Economic Scale (ES) serving as the key predictors in the analysis, and offers a data-driven understanding of how cities gain attention in the digital era (\autoref{mt}). TQ represents the intrinsic appeal of tourism by aggregating the quality and quantity of cultural and natural attractions within a city. TC is based on a gravity model \cite{isard1954location} that includes the quality of scenic areas, distances between cities, and city population, capturing how systemic connectivity contributes to visibility beyond the availability of local resource. TP quantifies physical accessibility through the presence of airports and high-speed rail links to indicate a city integration into national mobility systems. ES serves as a proxy for urban capacity and prominence, measured by the natural logarithm of gross domestic product and rescaled to allow for meaningful comparison.

We employ a Poisson regression model to examine the mention count as the dependent variable, a non-negative integer that reflects how frequently a city appears in textual data and thereby captures the intensity of human attention it receives. We specify the model as
\begin{equation}
    \log \mathbb{E}(M_i) = \beta_0 + \beta_1 \mathrm{TC}_i^* + \beta_2 \mathrm{ES}_i^* + \beta_3 \mathrm{TP}_i^* + \beta_4 \mathrm{TQ}_i^*,
\end{equation}
where $M_i$ represents the observed number of mentions for the city $i$, and $\mathbb{E}(M_i)$ denotes its expected value. Each predictor on the right-hand side of our equation is rescaled to the interval $[0,\,1]$ via MinMax normalization, $X^* = (X - X_{\min}) / (X_{\max} - X_{\min})$, enabling direct comparison of their effects. Modelling the log-linear relationship between predictors and the expected count reveals how urban characteristics shape the distribution of human attention across cities.

As depicted in Table \ref{tab:poisson_us_cn}, our analysis reveals several critical insights into the determinants of user attention towards cities. Firstly, both ES and TP, which reflect GDP and geographic location-related transport infrastructure, respectively, exhibit strong positive correlations with city mention counts. Specifically, for the United States, ES has a coefficient of $2.832$ ($SE = 0.153$, $p < 0.001$), while TP shows a coefficient of $0.287$ ($SE = 0.026$, $p < 0.001$). Similarly, in China, ES and TP demonstrate coefficients of $6.188$ ($SE = 0.359$, $p < 0.001$) and $0.484$ ($SE = 0.070$, $p < 0.001$), respectively. These findings underscore the pivotal role of economic strength and accessibility in attracting visitor interest, suggesting that robust economies and well-connected transportation networks significantly enhance user engagement.

Moreover, TQ which measures famous scenic tourist areas, also exhibits a significant positive relationship with city mentions. In the United States, TQ has a coefficient of $12.106$ ($SE = 0.159$, $p < 0.001$), while in China, it is $5.794$ ($SE = 0.368$, $p < 0.001$). This aligns with conventional wisdom, indicating that high-quality tourism resources are indeed effective in drawing attention from potential visitors.

Counterintuitively, TC which integrates population, city distance and tourism quality, presents a significant inverse relationship with city mentions. The negative coefficients for TC are $-2.204$ ($SE = 0.230$, $p < 0.001$) in the United States and $-8.167$ ($SE = 0.639$, $p < 0.001$) in China. This counterintuitive finding suggests that merely combining rich tourism resources with high population density and geographical proximity does not necessarily translate into increased user attention. Instead, it implies that complex inter-city dynamics may dilute the effectiveness of these resources in capturing human attention.

\begin{table}[htbp]
\centering
\caption{Poisson regression results for city mention counts. Coefficients and their corresponding standard errors (SE) are reported for each predictor variable. In both the United States and China, the coefficients for Tourism Quality (TQ), Transportation Potential (TP) and Economic Scale (ES) are positive with p-values $<0.001$, highlighting significant impacts on city mention counts. However, the Travel Competitiveness (TC) shows a negative coefficient with P-values $<0.001$ in both countries, suggesting a significant inverse relationship between comprehensive tourism competitiveness (considering population, city distance, and tourism quality) and city mentions. The constant terms are also significant in both models.}
\label{tab:poisson_us_cn}
\begin{tabular*}{\textwidth}{@{\extracolsep{\fill}} l c c | c c }
\toprule
& \multicolumn{2}{c}{United States} & \multicolumn{2}{c}{China} \\
\cmidrule(lr){2-3} \cmidrule(lr){4-5}
Variables & Coefficient (Standard Error) & P-value & Coefficient (Standard Error) & P-value \\
\midrule
TQ & $12.106$ ($0.159$) & $<0.001$ & $6.196$ ($0.356$)  & $<0.001$ \\
TP  & $0.287$ ($0.026$)  & $<0.001$ & $0.572$ ($0.102$)  & $<0.001$ \\
TC        & $-2.204$ ($0.230$) & $<0.001$ & $-8.932$ ($0.619$) & $<0.001$ \\
ES & $2.832$ ($0.153$)  & $<0.001$ & $6.671$ ($0.343$)  & $<0.001$ \\
const           & $-5.407$ ($0.086$) & $<0.001$ & $0.194$ ($0.113$)  & $<0.001$ \\
\bottomrule
\end{tabular*}
\end{table}

These results consistently corroborate our initial hypothesis and the findings via complex networks, highlighting the intricate interplay between GDP, geographic factors, and tourism appeal. Economic strength and strategic geographic positioning, particularly through enhanced transportation infrastructure, play crucial roles in shaping collective attention. Meanwhile, the negative correlation associated with TC underscores the need for a more nuanced approach to tourism development, emphasizing the importance of balancing resource availability with effective utilization and accessibility.

\section{Conclusion and discussion}
Our analysis of large-scale collective attention data generated during periods of restricted human mobility reveals how economic scale, particularly GDP, becomes embedded in human attention of urban prominence across the United States and China --- two leading global economies in the Western and Eastern hemispheres. By constructing city-level attention networks and applying community detection and structural hole analysis grounded in complex network theory, we identify that regional agglomeration and functional coherence significantly shape the collective attention of urban systems. The integration of government-open data into a Poisson regression framework further quantifies the influence of tourism quality, transportation potential, economic scale, and a composite travel competitiveness index on collective urban attention. While GDP, transportation infrastructure, and tourism quality show significant positive associations with observed urban attention, consistent with our findings in attention networks, the travel competitiveness index, a composite measure based on the economic gravity model and incorporating population, iconic attractions and intercity distance, exhibits a significant negative association with human attention. This counterintuitive outcome suggests that collective attention responds more strongly to simple, cognitively accessible signals such as economic size, transport access, and iconic attractions than to system-level metrics of city appeal, highlighting a disconnect between policy priorities and public attention in the digital era.

The central contribution of this work lies in the integration of complex network science and computational econometrics into a unified analytical framework that bridges abstract network structures with measurable socioeconomic mechanisms. This synthesis advances the theoretical understanding of urban prominence by grounding cognitive representations of cities in empirical data on collective attention, offering a novel perspective on urban dynamics, human behavior and humanistic geography. Our findings offer practical guidance for urban communication and city branding, indicating that highlighting concrete and visible features, such as landmarks or cultural icons, connects more with the public than strategies based on composite competitiveness indices.

Several limitations must be acknowledged. The range of available socioeconomic indicators is constrained by data availability and scope of research themes, limiting the depth of explanatory power and leaving room for additional variables that may influence attention allocation. Although advanced AI-based algorithms have been developed for related projects, this investigation emphasizes methodological robustness through classical network and text analysis techniques, prioritizing reproducibility over algorithmic novelty. Our future research will incorporate multiple types of data, track how public attention to cities changes over time, and extend comparisons to other cultural contexts beyond the United States and China. Expanding the analysis to additional regions will allow more diverse cultural and urban contexts to be included. Improving measurement methods may support more reliable comparisons, provided that novel analytical tools eventually achieve sufficient robustness and reproducibility.

\section*{Data availability}\label{data}
The dataset used in this study is publicly available at https://github.com/wordbomb/econ-geo-city-attention-network.

\section*{Code availability}\label{code}
The source code is publicly available at https://github.com/wordbomb/econ-geo-city-attention-network.

\section*{Acknowledgements}
This work is supported by the National Natural Science Foundation of China (61803047), the Social Sciences Fund of Jiangsu Province 24XWB004, Ke-ke Shang is supported by Jiangsu Qing Lan Project. Michael Small is supported by .

\section*{Author contribution}
Ke-ke Shang conceived the study, supervised the research, designed the experiments and algorithms, performed coding and data analysis, drafted the initial manuscript, and critically reviewed the final version. Jiangli Zhu contributed to the experimental design and data collection and contributed equally to this work. Junfan Yi contributed equally to this work and participated in data collection, data cleaning, coding, data analysis, and visualization. Liwen Zhang contributed to the visualization and initial manuscript drafting. Junjie Yang provided critical review and revision of the manuscript. Ge Guo and Zixuan Jin contributed to data collection and cleaning. Michael Small contributed equally to this work, provided overall supervision, contributed to data analysis, and critically reviewed the manuscript. All authors read and approved the final manuscript.

\section*{Competing interests}
The authors declare no competing interests.

\section{Methods}\label{mt}
\subsection{Tourism Quality (TQ)}
Urban appeal originates from cultural and natural assets such as historic landmarks, museums, and scenic landscapes. Fortunately, official and commercial organizations provide standardized assessments of attraction quality. In China, national ratings evaluate sites based on expert reviews and visitor feedback. Similarly, major travel platforms use aggregated user reviews to gauge popularity and quality for the US. We capture this intrinsic appeal as the original tourism quality (OTQ), a simple index incorporating the number, type, and recognized ratings of key attractions:
\begin{equation}
\mathrm{OTQ}_i = \sum_{k=1}^{5} w_k \cdot N_k,
\label{eq:otq}
\end{equation}
where $N_k$ is the number of $k$-star attractions in city $i$. For the United States, weights are $w_5 = 10$, $w_4 = 8$, $w_3 = 4$, $w_2 = 2$, and $w_1 = 1$, based on aggregated TripAdvisor ratings. For China, only 3--5 star equivalents are included ($N_1 = N_2 = 0$), with weights $w_5 = 10$, $w_4 = 6$, and $w_3 = 2$, reflecting the national A-level attraction system; 1A and 2A sites are excluded as they are rarely recognized or promoted even by the attractions themselves due to their generally low quality.

To reduce right-skewness, stabilize variance, and enable elastic interpretation of regression coefficients, we apply the natural logarithm transformation to $\mathrm{OTQ}_i$ and define $\mathrm{TQ}_i = \ln(\mathrm{OTQ}_i)$. 
This approach is extended to all continuous independent variables in the analysis. 
To ensure the logarithm is well-defined even when values are zero, we add a small constant $\epsilon = 10^{-6}$ before transformation. 

\subsection{Travel Competitiveness (TC)}

Tourism quality provides a direct measure of the intrinsic appeal of attractions within a city. Inspired by the economic gravity model, we capture tourism potential through geographic proximity, where trip chaining enables multi-destination travel. The flow from city $j$ to city $i$ is proportional to the product of their tourism qualities and inversely proportional to the distance between them. Hence, we propose the tourism attraction score (TAS), a spatially integrated metric that quantifies the cumulative tourism potential attracted to city $i$ from all external cities within the defined radius.

\begin{equation}
\mathrm{TAS}_i = \sum_{j \neq i} \frac{\mathrm{OTQ}_i \cdot \mathrm{OTQ}_j}{d_{ij}}   
\end{equation}
where $\mathrm{OTQ}_i$ is the tourism quality of destination city $i$, $\mathrm{OTQ}_j$ is that of origin city $j$, and $d_{ij}$ is the great-circle distance between them. Recognizing that larger cities often exert disproportionately strong tourism attraction due to enhanced infrastructure, visibility, and centrality in city networks, we refine the Tourism Attraction Score via a multiplicative population scaling factor defined as the ratio of local population to the national mean.

\begin{equation}
\mathrm{TC}_i =  \ln(\mathrm{TAS}_i \cdot \frac{P_i}{\bar{P}}),
\end{equation}
where $P_i$ is the population of city $i$ and $\bar{P}$ is the mean population across all cities. This formulation emphasizes the synergy between regional connectivity and demographic scale. The resulting index, travel competitiveness (TC), measures systemic competitiveness, the strategic advantage derived from combining spatial integration with a large population base.

Crucially, TQ captures the intrinsic appeal of a city own attractions while TC measures the competitive advantage that comes from being close to other popular destinations and is scaled by population. Including both variables helps the model distinguish a city own tourism strengths from the advantages of its regional connections and avoids mixing local appeal with location-based benefits. 

\subsection{Transportation Potential (TP)}

Our complex network research reveals that in diverse geographical regions, information flows are most likely to pass through core cities, which are strategically positioned hubs surrounded by numerous neighboring urban centers. The hub cities benefit from prominent spatial locations that enhance connectivity, particularly enabled by airports and high-speed rail stations. As a result, geographical positioning shapes transportation access, which in turn determines visibility. Hence, we introduce transportation potential (TP), a composite measure that combines air and rail connectivity. 

\begin{equation}
\mathrm{TP}_i = \ln\left( w \cdot \mathbb{I}_\mathrm{airport} + (1 - w) \cdot \mathbb{I}_\mathrm{rail}  \right),
\end{equation}
where $\mathbb{I}_\mathrm{airport}$ and $\mathbb{I}_\mathrm{rail}$ indicate the presence of an airport and high-speed rail station of city $i$. The weight $w$ is $0.7$  for cities in the United States (air-dominated) and $0.5$ for Chinese cities, where high-speed rail complements air travel, reflecting a balanced multimodal system. . TP provides a binary, scalable measure of multimodal reach.

\subsection{Economic Scale (ES)}

Economic scale (ES) is often the most prominent characteristic influencing how cities are recognized and evaluated. GDP serves as a proxy for economic activity, encompassing production, trade, and public revenue, and reflects the resources available for transport systems, digital networks, and development. It drives investment in infrastructure and services, shaping urban connectivity and visibility. Here, we measure the ES of city $i$ as $\mathrm{ES}_i = \ln(\mathrm{GDP}_i)$, the natural logarithm of GDP, to address skewness and to enable the interpretation of regression coefficients in terms of both elasticity and semi-elasticity.

\begin{appendices}

\section{Extended Data}
\subsection{Attention Data}
\begin{figure}[h]
  \centering
  \includegraphics[width=1\textwidth]{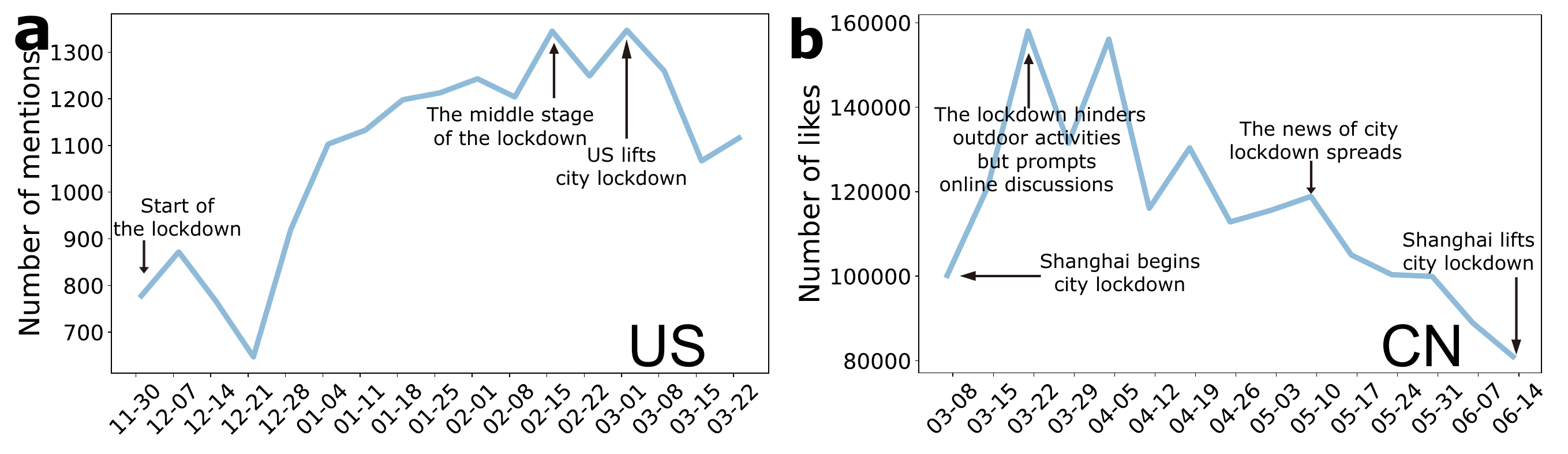} 
  \caption{Fluctuation in user engagement on travel platforms across the United States and China. During lockdown periods, public discussions about cities became increasingly concentrated in online forums, highlighting the Internet as a primary space for social interaction. In the US, mentions on \textit{TripAdvisor} peaked on February 15, 2022, during ongoing restrictions, and again on March 8, 2022, as cities began lifting lockdown. In China, likes on the top 20 weekly travel notes on \textit{Ctrip} surged from March to June 2022, particularly around the strict lockdown and reopening of Shanghai, reflecting strong virtual travel planning and increased interest in travel.}
  \label{fig:ml}
\end{figure}

All data are anonymized and publicly available. No personal or private user information was used. As depicted in \autoref{fig:ml}, in the US, open and public attention data were sourced from \href{https://www.tripadvisor.com}{TripAdvisor}, covering the period from November 30, 2021, to March 22, 2022. Travel discussions were collected from city forums. These data were compiled in July 2024 to analyze online travel discourse during periods of restricted mobility. For China, open and public attention data were obtained from \href{https://www.ctrip.com}{Ctrip}, from March to June 2022. Each week, the top 20 travel notes were collected. Ranked by user engagement, including likes, favorites, shares, and comments, over the past seven days. As only these top notes are shown on the homepage, the number of likes received by each city in these posts best captures overall public attention. This weekly collection ended after Ctrip's website redesign later around 2023.

\subsection{Network Data}
Following Louvain-based community detection, the US network comprised $396$ of $514$ cities ($1,865$ edges), with $118$ isolated; the Chinese network comprised $128$ of $183$ cities ($461$ edges), with $55$ isolated.

\begin{figure}[h]
    \centering
    \includegraphics[width=1\textwidth]{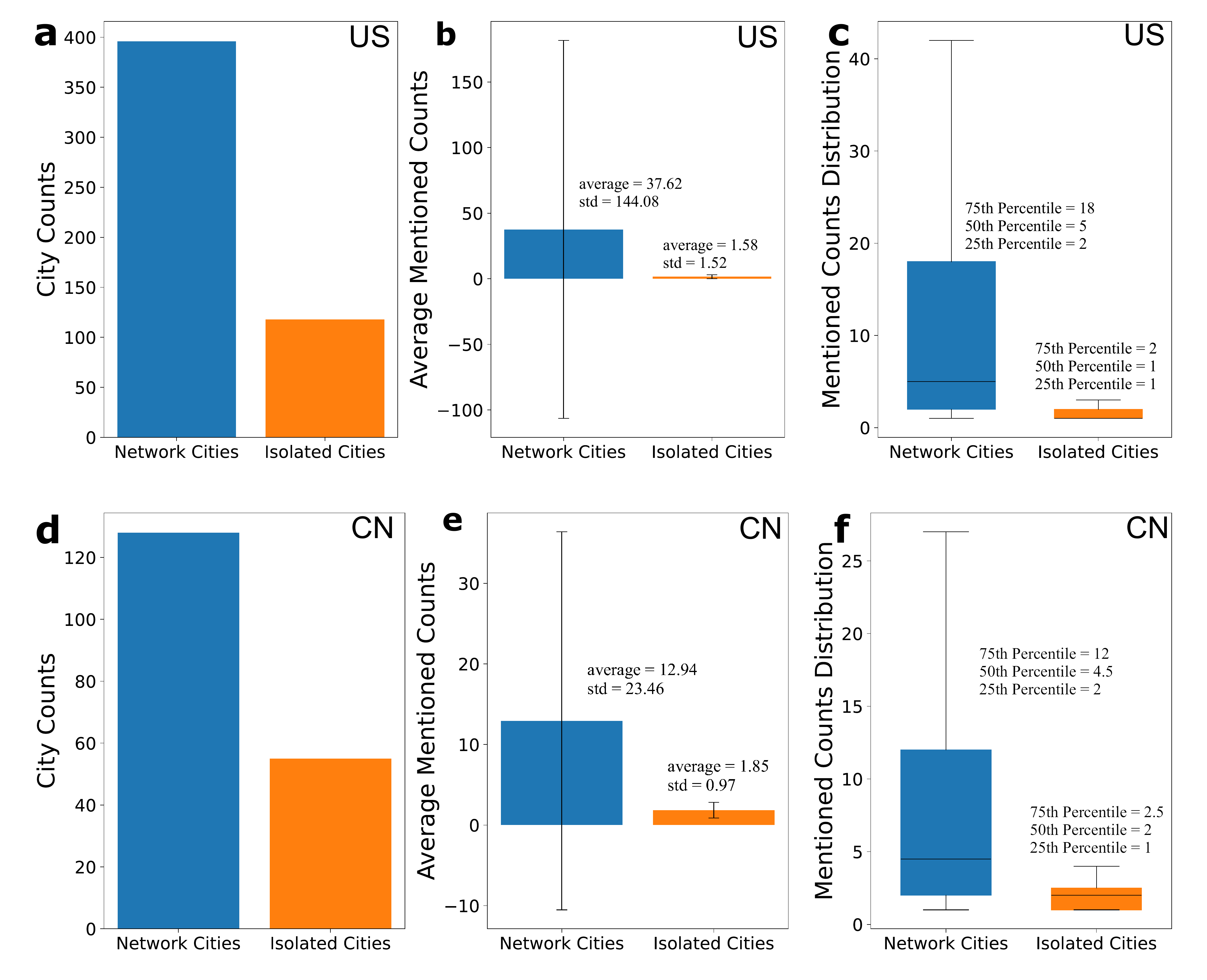}
    \caption{Comparison of network and isolated cities in the United States and China. Metrics include mean and standard deviation of mention counts (b), and distributional analysis via quartiles (c).}
    \label{cc}
\end{figure}

We evaluated the detection rationale by comparing the mention counts between the network and isolated cities in the United States and China (\autoref{cc}). Network cities in the United States exhibited a substantially higher mean mention count ($37.62$, SD = $144.08$) than isolated cities ($1.58$, SD = $1.52$). A consistent pattern emerged in China, where network cities had a mean mention count of $12.94$ (SD = $23.46$), far exceeding that of isolated cities ($1.85$, SD = $0.96$) (\autoref{cc} b and e).

Distributional analysis further confirmed this trend. The network cities in both countries exhibited substantially higher median, 25th, and 75th percentile mention counts (\autoref{cc} c and f), indicating consistently greater visibility. These results demonstrate that Louvain detection retains prominent, well-mentioned cities across diverse urban systems, while excluding those with minimal attention.

\subsection{Economic Data}
\begin{table}[htbp]
  \centering
  \caption{Data sources for the United States and China.}
  \label{data_sources}
  \begin{tabular}{|p{2cm}|p{4.7cm}|p{4.7cm}|}
    \hline
    \textbf{Data Type} & \textbf{United States} & \textbf{China} \\
    \hline
    Scenic Rating & 
    TripAdvisor (accessed July 2024) (\url{www.tripadvisor.com}) &
    Ministry of Culture and Tourism of PRC (\url{www.mct.gov.cn}) \\
    \hline
    Population & 
    U.S. Census Bureau (\url{www.census.gov}) &
    National Bureau of Statistics of PRC (\url{www.stats.gov.cn}) \\
    \hline
    GDP & 
    Bureau of Economic Analysis (BEA) (\url{www.bea.gov}) &
    National Bureau of Statistics of PRC (\url{www.stats.gov.cn}) \\
    \hline
    Railway & 
    OpenStreetMap (accessed July 2024) (\url{www.openstreetmap.org}) &
    12306 China Railway (accessed July 2022) (\url{www.12306.cn}) \\
    \hline
    Airport & 
    U.S. Department of Transportation (BTS) (\url{www.bts.gov}) &
    National Catalogue Service for Geographic Information (\url{www.webmap.cn}) \\
    \hline
  \end{tabular}
\end{table}

\subsection{Centrality and Mentions}
As shown in \autoref{cen}, the R-squared values for the centrality measures (Degree, Betweenness, and Closeness) are notably high, indicating that these metrics possess an exceptionally strong explanatory power for collective attention as measured by mentions rank. Specifically, the Degree Centrality rank in the United States yields an R-squared value of $0.700$, while Betweenness and Closeness centrality ranks achieve $0.600$ and $0.604$ respectively. In China, these values are also substantial, with R-squared values of $0.386$, $0.498$, and $0.459$. Furthermore, all p-values are less than $0.001$, demonstrating highly significant correlations. These results suggest that network centrality, particularly degree centrality, is a robust predictor of public attention and collective focus on cities. Consequently, cities with higher centrality can be reliably considered indicative of those receiving greater public attention, highlighting the critical role of network structure in understanding collective behavior.

\begin{table}[htbp]
\centering
\begin{tabular}{|c|c|c|c|}
\hline
\textbf{Country} & \textbf{Degree Centrality} & \textbf{Betweenness Centrality} & \textbf{Closeness Centrality} \\
\hline
United States 
& $R^2 = 0.700, p < 0.001$ & $R^2 = 0.600, p < 0.001$ & $R^2 = 0.604, p < 0.001$ \\
\hline
China 
& $R^2 = 0.386, p < 0.001$ & $R^2 = 0.498, p < 0.001$ & $R^2 = 0.459, p < 0.001$ \\
\hline
\end{tabular}
\caption{Linear Regression of Centrality Measures Rank on Mention Rank for Cities in the United States and China.}
\label{cen}
\end{table}

\section{Supplementary Information}
\subsection{Community detection for attention networks}
Modularity (Q value) \cite{newman2004finding} is the default metric for evaluating community detection algorithms. A higher Q value indicates that the detected communities have stronger internal connections, suggesting a better performance of the community detection algorithm, and a clear community structure will appear when the Q value is greater than $0.3$ \cite{clauset2004finding}. Under this metric, we have tested four fundamental types of benchmarks that are commonly used in the field of complex networks and also prevalent in social science research: the label propagation algorithm (LPA)\cite{raghavan2007near}
the Laplacian dynamics algorithm (LD)\cite{lambiotte2014random}, the Louvain algorithm\cite{blondel2008fast}, and the Girvan-Newman algorithm (GN)\cite{girvan2002community}. 

\begin{figure}[h]
  \centering
  \includegraphics[width=1\textwidth]{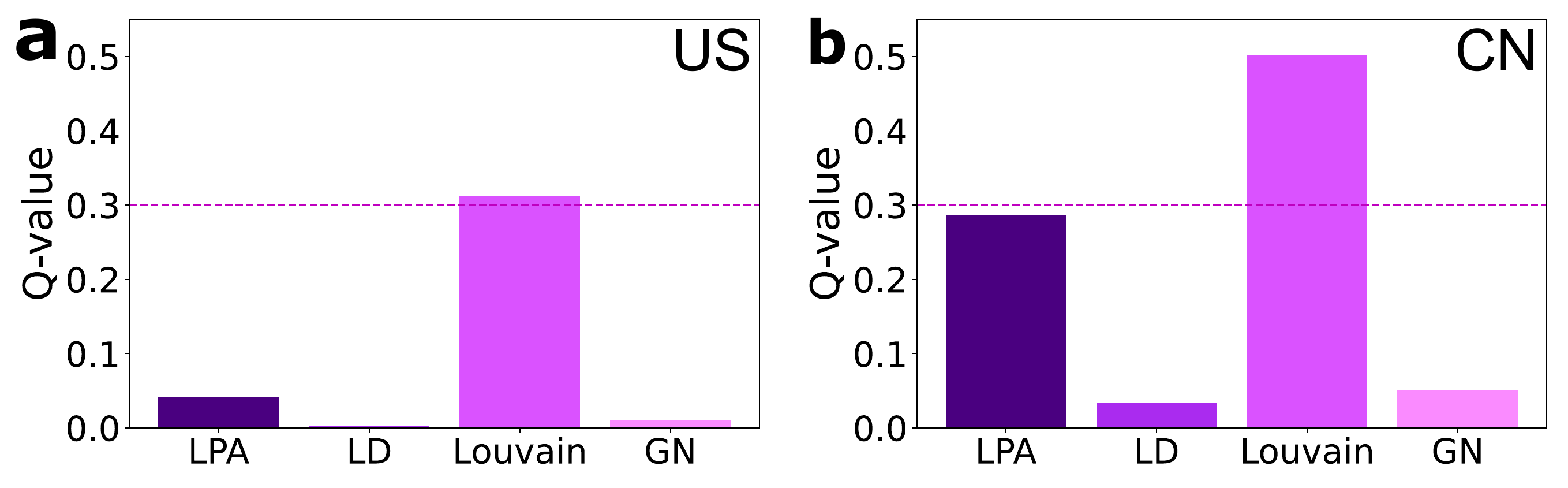} \hfill
  \caption{The Q-values of four commonly used community detection algorithms in social science, with $0.3$ typically serving as a benchmark for evaluating the success of community detection.}
  \label{cda}
\end{figure}

As depicted in \autoref{cda}, it is clear that the Louvain achieves the highest Q value. We can further see that the Q values of LPA, LD and GN are all lower than the threshold of $0.3$, although this method is widely recognized as being able to effectively identify community structures in many cases. In our experiments, the Louvain significantly outperformed them in terms of modularity, indicating its potential in discovering the structure of our city attention network.

\subsection{K-clique Percolation Method}
The K-clique method (KPM) is an algorithm for detecting communities within complex networks based on the concept of cliques\cite{Palla2005UncoveringTO}. A clique is a subset of nodes (representing cities or urban areas in this paper) in which all nodes are connected to each other, hence in our city networks, the clique will reveal the most tightly linked cluster of these cities or urban areas in terms of human attention or a specific relationship network. KPM identifies all the $k$-cliques in a network, where a $k$-clique is a clique consisting of $k$ nodes, and then groups these $k$-cliques into communities if they share a sufficient number of nodes. 

\begin{figure}[h]
  \centering
  \includegraphics[width=1\textwidth]{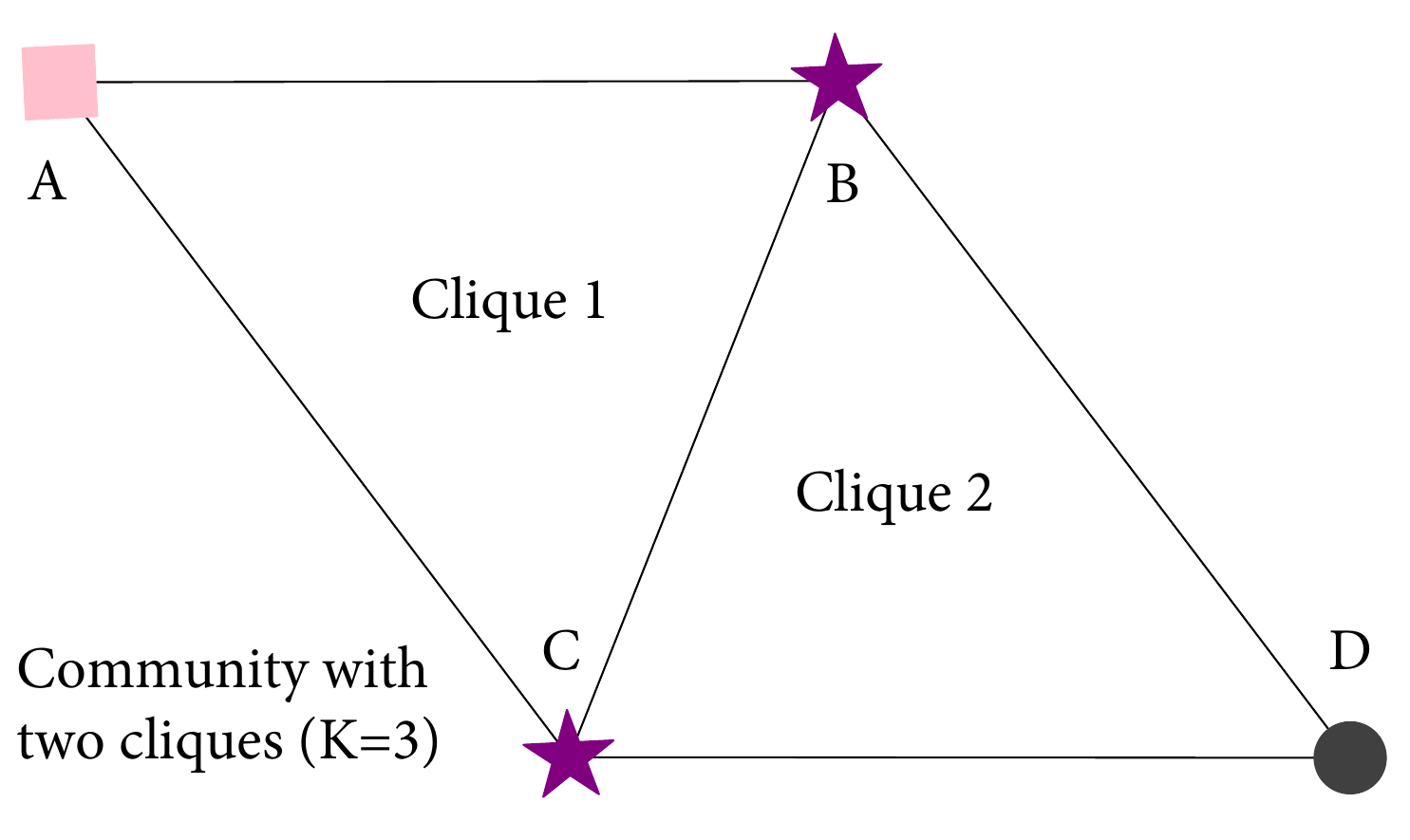} \hfill
  \caption{A simple network with four nodes labeled A, B, C and D. Suppose we are looking for 3-clique communities. Three nodes A, B, and C form the clique 1 (complete subgraph), and three nodes B, C, and D form the clique 2. Since cliques (A,B,C) and (B,C,D) share two nodes (B and C), they will be grouped together into one community.}
  \label{kpm}
\end{figure}

A toy model is shown in \autoref{kpm}, which proceeds in three steps. First, all $k$-cliques are detected by identifying node subsets that induce complete subgraphs of size $k$, where every pair of nodes is connected. Second, communities are formed by merging $k$-cliques that share at least $k-1$ nodes, based on the principle that extensive overlap indicates shared community membership. Third, the algorithm outputs a collection of communities, each composed of densely connected nodes.

\subsection{Topic coherence}
We establish LDA topic models for user comments based on community divisions\cite{blei2003latent} and verified the rationality of the divisions through topic coherence testing.  We adopt the coherence value (CV) \cite{mimno2011optimizing}, a leading coherence measure that evaluates the quality of topics by quantifying the consistency of the topics extracted from different portions of the same corpus. This method is widely utilized due to its effectiveness in assessing how reliably and uniformly themes can be identified across various splits of textual data.

Let \( F(w) \) denote the document frequency of the word type \( w \) (that is, the number of documents that contain at least one instance of the word type \( w \)), and let \( F(w, w') \) represent the co-document frequency of the word types \( w \) and \( w' \) (that is, the number of documents that contain one or more instances of both word types \( w \) and \( w' \)). We can define the coherence of a topic as follows:

\begin{equation}
CV(T;W(T)) = \sum_{z=2}^{Z} \sum_{y=1}^{z-1} \log\left(\frac{F(w_z, w_y) + \epsilon}{F(w_z)}\right),
\end{equation}
where \( W(T) = (w_1, w_2, \ldots, w_Z) \) is a list of the \( Z \) most probable words within the topic \( T \). A smoothing constant of $\epsilon$ is typically set to $1$ to prevent taking the logarithm of zero.

\subsection{Quantifying structure hole}
Our goal is to locate the influential cities across various communities embedded in attention networks. Structural holes refer to the gaps between pairs of nodes in a network that are not directly connected, and nodes that possess these gaps can act as intermediaries or bridges, thus acquiring informational advantages and control. The larger the scope of influence a gap node has and the fewer critical links it predictably controls, the more important it becomes. The effective size not only takes into account the size of a node ego network but also considers the redundancy of communication within the ego network. Here, centered on any node (the ego) in a graph, an ego network consists of the subgraph that includes that node and all of its directly connected neighbor nodes. This provides a localized view for analyzing the network structure and the connectivity between nodes. In other words, using this metric to measure structural holes takes into account both the influence of a node and the redundant connections in its communications. Therefore, this paper selects the effective size as the metric. 

\begin{equation}\label{ef}
\mathrm{ES}_i = N_i - \frac{2L_i}{N_i},
\end{equation}
where $N_i$ denotes the number of neighbors of node $i$, and $L_i$ is the number of edges among these neighbors. The term $N_i$ reflects the scope of direct influence, while $\frac{2L_i}{N_i}$ quantifies the average redundancy in the connections within the neighborhood. A higher $\mathrm{ES}_i$ indicates a more diverse and less cohesive local structure, signifying a greater potential for bridging disparate parts of the network.

\end{appendices}

\end{document}